\documentclass[preprint,showpacs,preprintnumbers,amsmath,amssymb]{revtex4}

\usepackage{graphicx}%
\usepackage{dcolumn}%
\usepackage{bm}%
\newcommand{\ia }{\'{\i}}

\begin{document}

\preprint{APS/123-QED}

\title{General segregation and chemical ordering in bimetallic
nanoclusters through atomistic view points}
\author{Juan Andr\'es Reyes-Nava$^{1}$}
\email{jareyesn@upchiapas.edu.mx}
\author{Jos\'e Luis Rodr\ia guez-L\'opez$^{2}$}
\email{jlrdz@ipicyt.edu.mx}
\author{Umapada Pal$^{3}$}
\email{upal@sirio.ifuap.buap.mx}
\affiliation{$^1$Universidad Polit\'ecnica de Chiapas,  Tuxtla
  Guti\'errez, Chiapas,  M\'exico} 
\affiliation{$^2$Advanced Materials Department, IPICYT;
San Luis Potos\'{\i}, S.L.P.,   M\'exico}
\affiliation{$^3$Instituto de F\'{\i}sica, Universidad
Aut\'onoma de Puebla, Puebla, Pue., M\'exico} 

\date{\today}

\begin{abstract}
 We predict general trends for surface segregation
in a binary metal cluster based on the difference
between the
atomic properties of the constituent elements.
Considering the attractive and repulsive
contributions of the cohesive energy of an atom
in a cluster, energetically
most favorable sites for a guest atom on a pure
metal cluster is determined. It is predicted that
for adjacent elements  in a row of the periodic
table, the bimetallic system would be more stable
if the component with smallest valence electron
density is placed on the surface. On the contrary,
for well separated components in the periodic
table,
the bimetallic cluster would be more stable if the
component with the smallest core electron density
is placed inside. Such chemical {\bf ordering trends}
in the
lowest energy configurations of Pt-Au, Pt-Pd and
Pt-Ni binary alloy clusters are verified for
their 561 atom systems through simulated annealing
process. It is predicted that the Ir, Rh, Ni, Pd
atoms would tend to be located inside the Ir-Pt,
Rh-Pd, Ni-Cu and Pd-Ag bimetallic nanoclusters,
respectively.
\end{abstract}


\maketitle

Surface segregation mechanism
in nanoalloy clusters is of great importance for
controlling their morphology, composition and
catalytic activities. Though, current
understanding of this phenomenon, based on some
theoretical calculations \cite{Friedel76,
Chris01,Ruban99,
Alden94,
Chelik84,
Mukh87,Bozzolo92,Ceder06,
sizeMismatched1,
Hwang}
and discrete experimental observations
\cite{Yang06,Rousset01,Vojislav07a,
AuPdStructuralIncoherency}
highlights the general trend of this process in
some particular systems, there exist no
generalized convention and physical explanation
to suit for all the binary systems till now.

Understanding of surface segregation in alloy
systems based on the properties of component
elements and composition has been addressed in
the pioneer works of Friedel \cite{Friedel76}.
For bulk alloys, several groups \cite{Chris01,
Ruban99,Alden94,Chelik84,Mukh87}
at different theory levels have confirmed the
otherwise intuitively general physical trend that
surface segregation energy in a transition metal
(TM) alloy is given by the difference in the
surface energies of the pure alloy components
\cite{Bozzolo92}. The present understanding of
this phenomenon is based on empirical and
thermodynamic models, although there have been
attempts to predict both the surface alloy
extent \cite{Chris01,Ruban99} and crystalline
structure of an alloy from quantum mechanical
perspectives \cite{Ceder06}.

Motivated by practical applications in
heterogeneous catalysis, surface segregation
process has also been addressed in metallic
nanoparticles (NPs). Baletto and coworkers
\cite{sizeMismatched1}
have explored the idea of atomic size mismatch
between the components to shed light on the
core-shell structure formation of different TM
alloys, where the size mismatch between the atoms
is one of the driving forces for the formation of
such structures. On the other hand Bozzolo
{\em et al.} \cite{Bozzolo92} have adapted a
simple semiempirical method based on the concept
that energy of formation of a given atomic
configuration is the sum of strain and chemical
energies of the individual atoms in the cluster,
to calculate the heat of formation of binary
alloys. However, though these criteria highlight
the general trends of surface segregation, they
do not give any physical explanation of the
phenomenon. Alternatively, Hwang {\em et al.}
have implemented a general methodology for the
quantitative determination of the extent of
alloying or atomic distribution in bimetallic
NPs, based on local coordination parameters
extracted from X- ray absorption spectroscopy
(XAS) data \cite{Hwang}. Although, the method
provides a good
qualitative description of the experimental
results, it is applicable to bimetallic NPs
prepared under certain conditions, which have
not fully reached their thermodynamic equilibrium.

In general, the cohesive energy of an atom in a
cluster is the sum of the energies corresponding
to its attractive and repulsive interactions with
the other atoms of the system:
$U_{coh}=U_{atr}+U_{rep}$ \cite{RepulsiveF}.
Also, the interaction of an atom with the others
of the system is mainly determined by its nearest
neighbors. Thus, the magnitude (absolute value)
of the changes $\Delta U_{atr}$ and
$\Delta U_{rep }$ at a site of a pure cluster
induced by the replacement of its original atom
with a guest one (guest-replacement),
fundamentally depends on the coordination number
of the site. The magnitudes of both the changes
reach their minima when the guest-replacement is
done at a cluster surface and maxima when the
replacement site is inside the cluster.
Because the values of $U_{atr}$ and $U_{rep}$
are negative and positive respectively, the most
energetically favored site for the
guest-replacement process is the replacement site
in which both $\Delta U_{atr}$ and
$\Delta U_{rep }$ reach their minimum values.

Now, the cohesive energy can be changed only by
three ways: {\em i}) only one of its contributions
changes, {\em ii}) both of them increase or
decrease, and {\em iii}) one contribution
increases and the other decreases. According to
the above described relation between these energy
changes and the coordination number, in the cases
{\em i}) and {\em ii}) the best replacement site
for the change of one contribution is also the
best for the change of the other. On the contrary,
for the situation {\em iii}) the change in
attractive contribution reaches its minimum value
at the same site at which the repulsive one
reaches its maximum. The best site for the change
of one contribution is the worst for the change of
the other. Therefore, while the most energetically
favored replacement site by the guest-host
replacement process can be {\em a priori}
determined in
cases {\em i}) and {\em ii}), it is not possible
in the case {\em iii}) and the changes in the
cohesive energy contributions must be calculated.

In addition, the differences between the
properties of the guest atom and those of the
host atoms define the values of $\Delta U_{atr}$
and $\Delta U_{rep}$ at the replacement site. In
a first approximation, the core and the valence
electron charge densities of the atoms determine
the extent of their repulsive and attractive
interactions, respectively.
If we assume that the electron
charge transfer between the atoms inside the
cluster is practically negligible, the extent of
the energy changes induced at a given site can be
{\em a priori} determined only from the core and
valence electron densities of the neutral host
and guest atoms. Independently, each of the two
electron charge densities of the guest atom can
be less, equal or higher than the corresponding
one of the host atom. When the guest and the host
atoms have adjacent locations in a row on the
periodic table, their core electron charge
densities are similar, thus
$\Delta U_{rep}\approx0$, and due to the
difference between their valence charge
densities, $\Delta U_{atr} \neq 0$. Thus, this
situation belongs to case {\em i}). The more
energetically favored site for the
guest-replacement will be a surface site if the
valence electron density of the guest atom is
lower, and it will be a core site if the valence
electron density is the higher. On the contrary,
if the guest and the host atoms are far enough
away on the periodic table, the difference
between their core electron charge densities will
be more important than the difference between
their valence electron charge densities. Then,
even in case {\em iii}), the more energetically
favored site for the replacement will be a
surface site if the core electron density of the
guest atom is the higher one, and will be a
volume site if it is the smaller one.

Therefore, with respect to the relative location
of the component atoms in the periodic table,
there are two situations at which the more stable
chemical ordering in a binary metal cluster alloy
can be {\em a priori} determined. {\em When they
have adjacent locations in a row, the component
with smaller valence electron density will be on
the surface}. On the contrary, {\em if the alloy
components have atomic numbers far enough away,
the metallic atom with smaller core electron
density will be in the core}. As immediate
consequence of this result, stable Pt-Au clusters
can be obtained by putting the gold atom on the
surface, and the tendency of the Ni atoms to be
located in the core is higher in the Pt-Ni alloy
than for Pd atoms in the Pt-Pd alloy.

\begin{figure}
\includegraphics[scale=0.36,angle=-90]{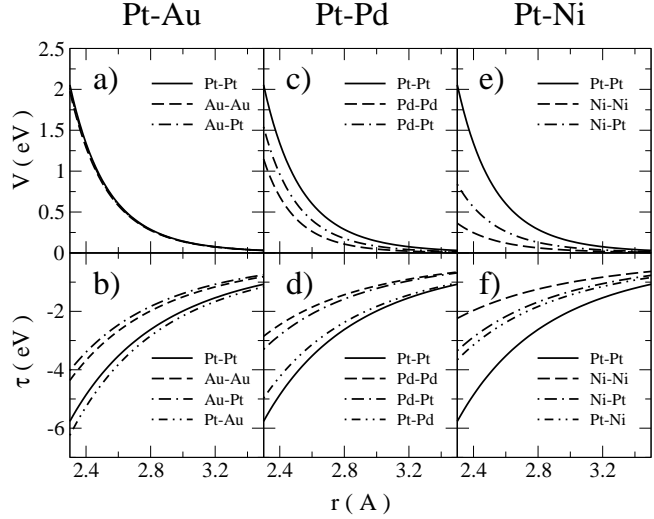}
\caption{\label{fig:potentialpieces}
Interaction energies
needed to make up the model potential for each
binary system. For the $ij$-th pair of atoms,
$V^{ij}(r)$ is the repulsive potential energy.
The function $\tau^{ij}(r)$ represents the
attractive energy of the $i$-th atom in the
presence of the $j$-th atom.}
\end{figure}

In order to support these predicted results, the
chemical ordering in Pt-Au, Pt-Pd and Pt-Ni alloy
clusters of 561 atoms are investigated. The
atomic interaction in the alloy nanoclusters is
modelled by the many body Rafii-Tabar \& Sutton
version of the Sutton \& Chen potential that is
based on the second moment approximation of a
tight binding Hamiltonian
\cite{BiMetRafiiSutton}. In this model, the
cohesive energy of the $i$-th atom is constructed
from the repulsive $V^{ij}(r)$ and the attractive
$\tau^{ij}(r)$ energies of its independent
interactions with the $j$-th atoms ($j \ne i$).

Atomic interactions in the alloy cluster are
completely determined by these functions. Both
the sets of functions depend on the nature of
the $i$-th and $j$-th atoms. The $\tau^{ij}(r)$
functions are explicitly given by
$\tau^{AA}(r) = -d^{AA} \left\{\phi^{AA}(r) \right\}^\frac{1}{2}$,
$\tau^{AB}(r) = -d^{AA} \left\{\phi^{AB}(r) \right\}^\frac{1}{2}$,
$\tau^{BA}(r) = -d^{BB} \left\{\phi^{AB}(r) \right\}^\frac{1}{2}$,
$\tau^{BB}(r) = -d^{BB} \left\{\phi^{BB}(r) \right\}^\frac{1}{2}$.
The superscripts specify the kind of the $i$-th
and $j$-th atoms, and the $\phi^{ij}(r)$ and
$V^{ij}(r)$ terms as well as the constants
$d$'s are defined in Ref. \cite{BiMetRafiiSutton}.

The sets of attractive and repulsive functions
for each binary alloy clusters are displayed in
the Fig. \ref{fig:potentialpieces}. For the
$ij$-th atom pair, the attractive energy of the
$i$-th atom due to the presence of the $j$-th
atom is given by the $\tau^{ij}$ term. The
repulsive energy of any one atom of the pair is
given by the function $V^{ij}$. For the atom
pairs Au-Pt, Pd-Pt and Ni-Pt, the function
$\tau^{ij}$ (taking Pt atom as $j$-th atom) is
higher than for the atom pair Pt-Pt (see dotted
and continuous curves in b), d) and f) sections
of the Fig. \ref{fig:potentialpieces}), as an
expression of the fact that the valence electron
densities of the Au, Pd and Ni atoms are less
than that of the Pt atom. On the other hand, the
repulsive energy $V^{ij}$ of the Pt-Pt atom pair
is almost equals that of the Au-Pt, and greater
than those of Pd-Pt and Ni-Pt atom pairs (see
continuous line and dotted curves of the Figs.
\ref{fig:potentialpieces}a-\ref{fig:potentialpieces}c).
The results are expected, as the core electron
density of Pt is higher than those of Pd and
Ni, and almost the same as that of Au.

\begin{figure}
\includegraphics[scale=0.36,angle=-90]{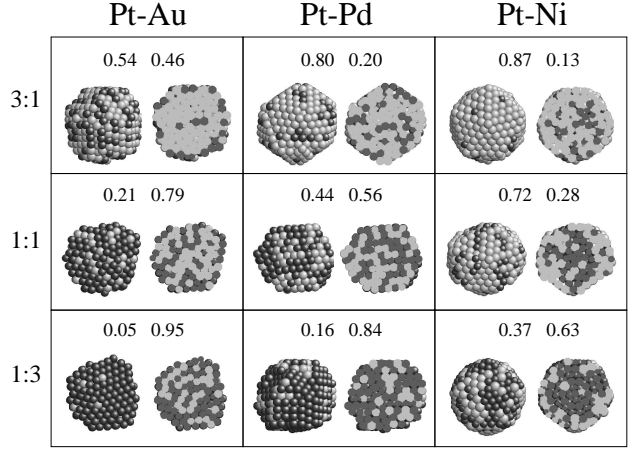}
\caption{\label{fig:chemicalorder}
Candidates for the lowest
energy structures of each binary system
(columns), at the 3:1, 1:1 and 1:3 global
compositions (rows). The superficial and cross
sectional views of a particular cluster are shown
in each block. Lighter spheres represent platinum
atoms. The values presented in each block are the
superficial compositions of the clusters.}
\end{figure}

Candidates for the lowest energy structure were
found by means of constant energy MD simulations
using the model potential. The clusters from five
dynamic states (amorphous structure with uniform
specie distributions) periodically extracted from
a long simulation ($\sim$ 0.1 $\mu$s) of the
liquid phase were frozen at a rate of 10 K/ns.
The final configurations obtained from this
simulated annealing process are described in Fig.
\ref{fig:chemicalorder}. Although various
morphologies were found like icosahedral, Mark's
decahedral and cuboctahedral, their species
distributions are practically the same for the
five homotops. For a uniform species
distribution, the surface and global (the system
as a whole) compositions are equal. Thus,
according to the theoretical results, the values
of the superficial composition for Au in Pt-Au
must be greater than the global composition
value. In addition, the superficial concentration
of Ni in Pt-Ni must be less than that of Pd in
Pt-Pd. The superficial composition values of Au,
Pd and Ni for the most stable structures of
Pt-Au, Pt-Pd, and Pt-Ni alloy nanoclusters were
found to confirm the predicted results. For the
Pt-Au clusters, the values of the superficial
composition of Au: 0.46, 0.79 and 0.95, are
greater than their corresponding global
concentration values: 0.25, 0.50 and 0.75 (see
the values in the Fig. \ref{fig:chemicalorder}).
In addition, for the Pt-Ni system the superficial
composition values of Ni: 0.13, 0.28 and 0.63,
are less than the ones corresponding to Pd in
Pt-Pd system: 0.20, 0.56 and 0.84 (see the values
in the second and third columns of Fig.
\ref{fig:chemicalorder}). Thus, the theoretical
predictions are completely supported by the model
potential used to describe these three nanoalloy
systems. However, the final proof of our
predictions comes from the reported experimental
results. Recently, using several experimental
techniques, Yang and coworkers have shown that
Pt-Au nanoparticles with core-shell structure can
be obtained only when nanoparticles of Pt are
used as seeds. On the contrary, only monometallic
nanoparticles of each constituent metal are
produced when the synthesis order is reversed
using Au nanoparticles as seeds \cite{Yang06}.
Though there are experimental evidences of
existing stable Pt-Pd clusters with Pd enriched
surfaces \cite{Rousset01} and also the stable
Pt-Ni alloys with Pt enriched extended surface
\cite{Vojislav07a}, the nanostructures were
grown through low temperature synthesis
processes, where a complete thermodynamic
equilibrium is not achieved. The experimental and
theoretical evidences for the higher stability of
Pd(core)-Au(shell) reported by Pal's group
\cite{AuPdStructuralIncoherency} provide additional
validity of the present results. The Pd-Au system
falls under the case {\em iii}). Thus, Pd with
smaller core electron density will be in the
volume. The two components of Ir-Pt, Rh-Pd, Ni-Cu
and Pd-Ag clusters have adjacent locations in a
row of the periodic table, and the earlier ones
have the higher valence electron charge density
for each system. Thus, it can be predicted that
the earlier elements of the binary clusters would
tend to be located inside the bimetallic cluster;
just as it occurs for Pt-Au system.

\begin{figure}
\includegraphics[scale=0.32,angle=-90]{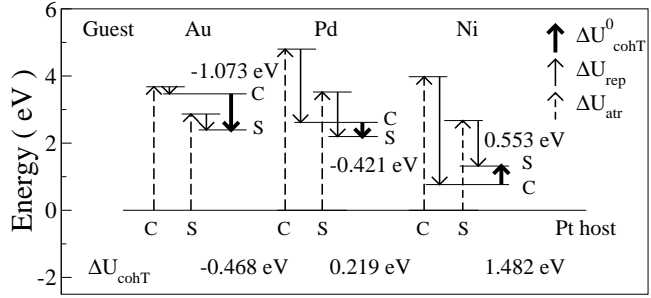}
\caption{\label{fig:hostPtcambioe}
Energy changes at a
site of a Pt cluster, induced by the replacement
of its original atom with a guest one. The
changes are calculated at the center ($C$) and a
surface ($S$) site. The horizontal line
represents the original energy of the site, the
dashed arrow corresponds to $\Delta U_{atr}$ and
the following continuous arrow to
$\Delta U_{rep}$. The term $\Delta U^{0}_{cohT}$
(see the text) is represented by the bold arrow.
Note that the sign of this term is the same as
that of $\Delta U_{cohT}$ (see the text) for
Pt-Au and Pt-Ni.}
\end{figure}

A quantitative description of the analysis
presented here can be given through the energy
changes shown in the Fig. \ref{fig:hostPtcambioe}.
The change $\Delta U_{cohT}$ represents the
difference in cohesive energy of the surface
homotop (with a guest atom at the surface) of an
icosahedral Pt cluster with respect to its
central homotop (with the guest atom at the
volume). According to the sign of
$\Delta U_{cohT}$, the most stable homotop is the
central one for Pd and Ni guest atoms, and the
surface one for Au. This is why the superficial
compositions of Pd and Ni atoms are less than
0.25, and that of the Au is higher than 0.25 for
the Pt-Pd, Pt-Ni and Pt-Au nanoclusters with
global atomic compositions 3:1, respectively (see
the values presented in the first row of Fig.
\ref{fig:chemicalorder}). The expression of
$\Delta U_{cohT}$ as the sum of the cohesive
energy changes at all the sites of the cluster;
the replacement sites, the nearest sites to the
replacement sites and so on, can be used to
determine the sign of $\Delta U_{cohT}$. The
first term of this sum is given by
$\Delta U^{0}_{cohT} = \Delta U_{coh}(S)-
\Delta U_{coh}(C)$, where $\Delta U_{coh}(i)$
is the change of the cohesive energy induced at
the replacement site $i$ through the
guest-replacement; $S$ and $C$ denote the surface
site and the center site, respectively.

\begin{figure}
\includegraphics[scale=0.32,angle=-90]{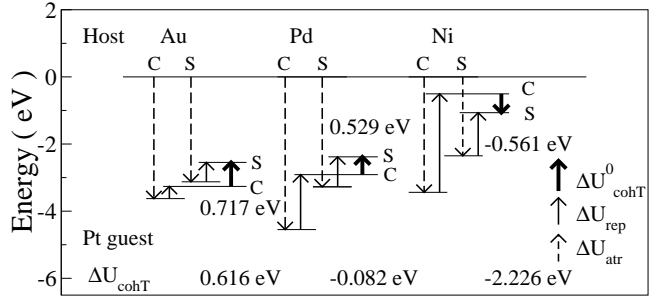}
\caption{\label{fig:guestPtcambioe}
Energy changes at a
site of a pure Au, Pd, Pt cluster induced by the
replacement of its original atom with a Pt guest
atom. The same quantities as in Fig. 3, but now
for three pure icosahedral hosts of 561 Au, Pd or
Ni atoms and a Pt guest atom. Again the sign of
$\Delta U^{0}_{cohT}$ is the same of
$\Delta U_{cohT}$ for Pt-Au and Pt-Ni.}
\end{figure}

The approximation $\Delta U_{cohT} \approx \Delta
U^{0}_{cohT}$ , is enough to obtain the sign of
$\Delta U_{cohT}$ for Au and Ni, but for Pd it is
insufficient (Compare the values of
$\Delta U_{cohT}$ at the bottom of the Fig.
\ref{fig:hostPtcambioe} with those of
$\Delta U^{0}_{cohT}$ at the top). Thus, for this
guest, the next term must be added. The
calculated changes of the repulsive and
attractive energy contributions as well as the
term $\Delta U^{0}_{coh}$ are represented in the
Fig. \ref{fig:hostPtcambioe}, by the continuous,
dashed and bold arrows, respectively.
The Fig. \ref{fig:guestPtcambioe} shows the same
quantities described in the Fig.
\ref{fig:hostPtcambioe}, but now for
three pure icosahedral hosts of 561 Au, Pd or Ni
atoms with a single Pt guest atom. This figure
shows that Pt is more stable at the center of the
Au and at surface of the Ni hosts. The fact that
$\Delta U_{cohT} \sim 0$ for Pd means that the
changes $\Delta U_{atrT}$ and $\Delta U_{repT}$
almost cancel each other for this situation of
case {\em iii}). Thus, for the most stable
chemical ordering of Pt-Pd nanoclusters, the
species distribution will be nearly homogeneous
and the surface and global compositions must be
similar (note that the superficial composition
values of the column Pt-Pd in the Fig.
\ref{fig:chemicalorder} are close to the global
composition values).

In summary, based on the fact that the cohesive
energy is the sum of its attractive and repulsive
contributions, it was shown that the trends of
the most stable chemical ordering in a binary
system are determined by the differences in the
atomic properties of its elemental components.
When they are adjacent in a row of the periodic
table, the component with lower valence electron
density will be on the surface. Moreover, if the
alloy components have atomic numbers enough
distant, the metallic atom with lower core
electron density will be in the volume. MD
simulations confirm the validity of these results
for Pt-Au, Pt-Pd, and Pt-Ni systems. Finally, it
is predicted that the Ir, Rh, Ni, Pd atoms would
try to be located inside the Ir-Pt, Rh-Pd, Ni-Cu
and Pd-Ag bimetallic nanoclusters, respectively.
We address that these predictions can be
validated for any bimetallic nanocluster, and
current research is underway.

\begin{acknowledgments}
This work was supported by {\sc conacyt}-M\'exico,
under the Project grants FOMIX CHIS-2006-C06-45675 and
J42645-F. The authors wish to thank Professors:
Karo Michaelian, Ignacio Garz\'on
({\sc IFUNAM, M\'exico}),  and Jos\'e Manuel Cabrera
({\sc FC-UASLP, M\'exico}) for helpful discussions,
and computational supports from the Texas Advanced
Computing Center ({\sc TACC-UT-Austin, USA}), and
the Centro Nacional de Superc\'omputo ({\sc
CNS-IPICYT, M\'exico}).
\end{acknowledgments}

\end{document}